\documentclass[sigchi-a, nonacm]{acmart}
\setcopyright{none}
\usepackage[type={CC}, modifier={by}, version={4.0}]{doclicense}
\usepackage{tikz}
\usetikzlibrary{arrows.meta, positioning, shapes.geometric, decorations.pathreplacing, fit}
\definecolor{ctp-blue}{HTML}{89b4fa}
\definecolor{ctp-green}{HTML}{a6e3a1}
\definecolor{ctp-yellow}{HTML}{f9e2af}
\definecolor{ctp-mauve}{HTML}{cba6f7}
\definecolor{ctp-peach}{HTML}{fab387}
\definecolor{ctp-red}{HTML}{f38ba8}
\definecolor{ctp-teal}{HTML}{94e2d5}
\definecolor{ctp-sky}{HTML}{89dceb}
\definecolor{ctp-lavender}{HTML}{b4befe}
\definecolor{ctp-text}{HTML}{cdd6f4}
\definecolor{ctp-subtext}{HTML}{a6adc8}
\definecolor{ctp-overlay}{HTML}{6c7086}
\definecolor{ctp-surface0}{HTML}{313244}
\definecolor{ctp-surface1}{HTML}{45475a}
\definecolor{ctp-base}{HTML}{1e1e2e}

\title{Terminal Is All You Need: Design Properties for Human-AI Agent Collaboration}

\author{Alexandre De Masi}
\affiliation{%
  \institution{University of Geneva}
  \country{Switzerland}}
\email{alexandre.demasi@unige.ch}

\begin{document}

\begin{teaserfigure}
\vspace{-1em}
\centering
\small
Accepted as poster at the CHI 2026 Workshop on Human-AI-UI Interactions Across Modalities (CUCHI'26), April 14, 2026, Barcelona, Spain.
\vspace{-0.5em}
\end{teaserfigure}

\begin{abstract}
While research on AI agents focuses on enabling them to operate graphical user interfaces, the most effective and widely adopted agent tools in practice are terminal-based. We argue that this convergence is not coincidental. It reflects three design properties central to effective human-AI-UI collaboration: representational compatibility between agent and interface, transparency of agent actions within the interaction medium, and low barriers to entry for human participants. We ground each property in established HCI theory, show how terminal-based tools satisfy them by default, and argue that any modality, including graphical and spatial interfaces, must be deliberately engineered to achieve them. Rather than a legacy artifact, the terminal serves as a design exemplar whose properties any agent-facing modality must replicate.
\end{abstract}

\maketitle

\begin{marginfigure}
\begin{tikzpicture}[
    term/.style={fill=ctp-base, text=ctp-text, rounded corners=3pt,
      minimum width=4.4cm, inner sep=4pt, font=\ttfamily\tiny, align=left, text width=4cm}
]
\node[term] (t) {%
\textcolor{ctp-green}{\$}\ \textcolor{ctp-text}{Fix the login bug}\\[2pt]
\textcolor{ctp-sky}{Agent:} Reading auth.py...\\
\textcolor{ctp-sky}{Agent:} Found missing None check\\
\textcolor{ctp-sky}{\ \ \ \ \ }  on line 42.\\
\textcolor{ctp-sky}{Agent:} Proposed fix:\\[2pt]
\textcolor{ctp-yellow}{--- app/auth.py}\\
\textcolor{ctp-green}{+\,if user is None:}\\
\textcolor{ctp-green}{+\ \ \ \ return None}\\[2pt]
\textcolor{ctp-mauve}{Apply?}\ \textcolor{ctp-green}{[y/n]}\ \textcolor{ctp-text}{y}\\[2pt]
\textcolor{ctp-green}{\checkmark}\ \textcolor{ctp-subtext}{File saved. Running tests...}\\
\textcolor{ctp-green}{\checkmark}\ \textcolor{ctp-subtext}{All 14 tests passed.}%
};
\end{tikzpicture}
\caption{Terminal-based agent interaction: natural language request, code diff, approval gate, and test execution in one text stream.}
\label{fig:terminal}
\end{marginfigure}

\section{Introduction}

AI agents that perceive, interpret, and operate user interfaces must serve a dual audience: the interface must remain legible and controllable by the human while being interpretable and operable by the agent. Yet current approaches to this dual-audience problem diverge sharply in their effectiveness.

GUI agents attempt to bridge this gap by perceiving screenshots or accessibility trees and generating clicks and keystrokes~\cite{xie2024osworld, gou2025uiworld}. The Agent-Computer Interface (ACI) framework treats this as a dedicated design problem, distinct from human-facing interface design~\cite{yang2024sweagent}. Yet the approach faces persistent difficulties: the best model achieves 12.24\% task success on OSWorld versus 72.36\% for humans~\cite{xie2024osworld}, and frontier web agents show up to 59\% lower success rates on realistic tasks than on simpler benchmarks~\cite{xue2025illusion}.

A different pattern is emerging among multi-step \textit{agentic} tools, which autonomously read files, plan changes, execute commands, and run tests. Unlike single-action tools like GitHub Copilot that suggest inline completions within graphical editors, agentic tools are converging on text-based sequential interaction (Figure~\ref{fig:terminal}). By late 2024, AI-generated code accounted for approximately 30\% of new Python functions on GitHub~\cite{daniotti2026who}. A study of over 129,000 GitHub projects found coding agent adoption rates between 15\% and 23\%, with rapid growth since early 2025 and traces of over 50 distinct agentic tools~\cite{robbes2026agentic}. Claude Code, OpenAI Codex, and Cursor's agent mode all center on a text-based sequential pattern, even though these tools are built by competing organizations. Cursor itself is a graphical IDE, yet its agentic capabilities operate through a terminal-like text channel embedded within the editor. Single-action code completion tools face their own adoption challenges: experienced open-source developers using AI assistants were measured as 19\% \textit{slower} than without AI, despite perceiving themselves as faster~\cite{becker2025measuring}. This perception-reality gap directly motivates the transparency property we identify below.

We do not take this as evidence that command-line interfaces (CLIs) are inherently superior. We take it as a signal that demands analysis. We identify three design properties that account for this convergence, ground them in established HCI theory, and argue that they constitute core design requirements for effective human-AI-UI collaboration. Modalities that fail to address them risk encountering the same difficulties that GUI agents face today.

\section{Background: The Dual-Audience Interface}

\begin{marginfigure}
\centering
\begin{tikzpicture}[
    actor/.style={draw=ctp-overlay, circle, minimum size=1.1cm, align=center, font=\tiny\bfseries, inner sep=1pt},
    prop/.style={font=\tiny, align=center, text width=2.2cm, fill=white, inner sep=1pt, rounded corners=1pt},
    edge/.style={thick, ctp-overlay}
]

\node[actor, fill=ctp-blue!20] (human) at (90:1.8) {Human};
\node[actor, fill=ctp-green!20] (agent) at (330:1.8) {Agent};
\node[actor, fill=ctp-yellow!25] (ui) at (210:1.8) {UI};

\draw[edge] (human) -- (agent);
\draw[edge] (agent) -- (ui);
\draw[edge] (ui) -- (human);

\node[prop, fill=ctp-mauve!12, draw=ctp-mauve!40, rounded corners=2pt]
  at (30:2.4) {\textbf{P2:} Transparency};
\node[prop, fill=ctp-mauve!12, draw=ctp-mauve!40, rounded corners=2pt]
  at (270:2.0) {\textbf{P1:} Representational Compatibility};
\node[prop, fill=ctp-mauve!12, draw=ctp-mauve!40, rounded corners=2pt]
  at (150:2.4) {\textbf{P3:} Low\\Barriers};

\end{tikzpicture}
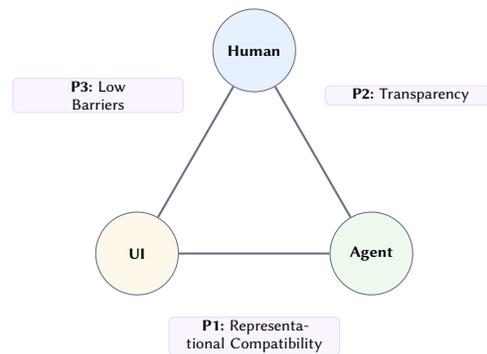
\caption{The Human-Agent-UI triangle with the three design properties mapped to its edges.}
\label{fig:properties}
\end{marginfigure}

Classical HCI optimized interfaces for a single audience: the human user. Direct manipulation reduces the cognitive burden of command recall and supports rapid visual feedback~\cite{shneiderman1983direct}. The Gulfs of Execution and Evaluation formalize these advantages: good interfaces minimize the distance between user intention and available action, and between system state and user perception~\cite{norman1986user}.

AI agents change this picture. In the Human-Agent-UI triangle (Figure~\ref{fig:properties}), the interface must support two kinds of interaction simultaneously. The human must observe, direct, and oversee the agent's work. The agent must perceive the interface state, select actions, and execute them reliably. These requirements can conflict. A GUI optimized for human visual perception may present the agent with a difficult grounding problem~\cite{gou2025uiworld, zhang2025api}. Conversely, an interface optimized for agent operation (such as a raw API) may achieve representational compatibility with the agent while remaining opaque to human oversight.

The design challenge is to identify interface properties that serve both audiences. Terminal-based agent tools offer a compelling case study: they have emerged organically from practice and demonstrably satisfy both requirements. Three properties account for this effectiveness.

\section{Property 1: Representational Compatibility}

The first property concerns the alignment between the representational format native to the agent and the format of the interface. When these match, the translation overhead between agent reasoning and interface action is minimized.

\begin{marginfigure}
\begin{tikzpicture}[
    term/.style={fill=ctp-base, text=ctp-text, rounded corners=3pt,
      minimum width=4.4cm, inner sep=4pt, font=\ttfamily\tiny, align=left, text width=4cm}
]
\node[term] (t) {%
\textcolor{ctp-subtext}{\# Agent output IS terminal input:}\\[2pt]
\textcolor{ctp-green}{\$}\ \textcolor{ctp-text}{refactor auth module}\\[2pt]
\textcolor{ctp-sky}{> mkdir app/auth}\\
\textcolor{ctp-sky}{> mv app/auth.py app/auth/core.py}\\
\textcolor{ctp-sky}{> cat app/auth/validate.py}\\[2pt]
\textcolor{ctp-green}{def validate(token: str) -> bool:}\\
\textcolor{ctp-green}{\ \ \ \ payload = jwt.decode(token)}\\
\textcolor{ctp-green}{\ \ \ \ return payload is not None}\\[2pt]
\textcolor{ctp-subtext}{\# Same medium, no translation.}%
};
\end{tikzpicture}
\caption{Agent text output maps directly to shell commands and Python code with no perception layer needed.}
\label{fig:repcompat}
\end{marginfigure}

Large language models process and generate text. Their mechanisms for interacting with external systems, from function calling to code generation to shell commands, are all text-based. Terminal interfaces likewise consume and produce text. The structural mapping is direct: tool invocations correspond to shell commands, parameters to arguments, and output arrives through standard text streams (Figure~\ref{fig:repcompat}). No additional perception or translation layer is required.

A carefully designed text-based ACI outperformed agents using the default Linux shell by 10.7 percentage points on SWE-bench~\cite{yang2024sweagent}. Using executable Python code as the agent action space yields up to 20\% higher success rates than JSON-based function calling, because code is a more expressive text-based representation~\cite{wang2024codeact}.

GUI agents, by contrast, must solve a perception problem before they can act: interpreting pixel-level screenshots, computing spatial coordinates, and mapping these to motor actions~\cite{gou2025uiworld}. This translation layer is the primary bottleneck for GUI agent performance~\cite{zhang2025api}.

\textbf{Implication for other modalities.} Representational compatibility alone is insufficient. Raw APIs are text-based but opaque to human oversight and inaccessible to non-expert users. The terminal succeeds because it achieves compatibility \textit{within a medium the human can also read}. For GUI agents, providing semantic interface representations (accessibility trees, DOM structures, UI element metadata) rather than raw pixels would reduce translation overhead while preserving human legibility.

\section{Property 2: Transparency of the Interaction Medium}

The second property concerns whether the interaction medium makes agent actions, reasoning, and history visible and inspectable by the human collaborator.

\begin{marginfigure}
\begin{tikzpicture}[
    node distance=0.35cm,
    box/.style={draw=ctp-overlay, rounded corners=2pt, minimum width=4.2cm, minimum height=0.55cm,
      align=center, text width=3.8cm, font=\tiny, inner sep=2pt},
    arr/.style={-{Stealth[length=1.5mm]}, semithick, ctp-overlay},
    lbl/.style={font=\tiny\bfseries, text=ctp-overlay, anchor=east}
]

\node[box, fill=ctp-blue!15] (intent) {\textbf{Human:} Natural language intent};
\node[box, fill=ctp-yellow!20, below=of intent] (request) {\textbf{Stream:} User request};
\node[box, fill=ctp-green!15, below=of request] (plan) {\textbf{Agent:} Interpret \& plan};
\node[box, fill=ctp-green!15, below=of plan] (exec) {\textbf{Agent:} Execute via UI};
\node[box, fill=ctp-green!15, below=of exec] (explain) {\textbf{Agent:} Explain \& report};
\node[box, fill=ctp-yellow!20, below=of explain] (output) {\textbf{Stream:} Agent output};
\node[box, fill=ctp-blue!15, below=of output] (review) {\textbf{Human:} Review \& approve};
\node[box, fill=ctp-yellow!20, below=of review] (log) {\textbf{Stream:} Session log};

\draw[arr] (intent) -- (request);
\draw[arr] (request) -- (plan);
\draw[arr] (plan) -- (exec);
\draw[arr] (exec) -- (explain);
\draw[arr] (explain) -- (output);
\draw[arr] (output) -- (review);
\draw[arr] (review) -- (log);

\draw[arr, ctp-mauve] (review.east) -- ++(0.25,0) |- (request.east);

\end{tikzpicture}
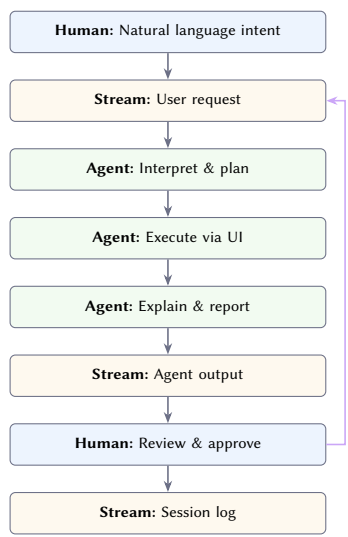
\caption{The interaction loop: the text stream (yellow) unifies communication, explanation, and audit.}
\label{fig:framework}
\end{marginfigure}

In terminal-based agent tools, the text stream simultaneously serves as communication channel, explanation surface, chronological record, and approval mechanism. Every proposed action is expressed as readable text. The sequential nature of terminal output produces a log of every decision point. Approval gates implement human-in-the-loop oversight with minimal additional instrumentation (Figure~\ref{fig:framework}).

Trust in automation depends on perceived competence, predictability, and the operator's ability to understand behavior~\cite{lee2004trust}. Transparency and user control are prerequisites for appropriate reliance~\cite{lee2004trust, shneiderman2022human}. Effective designs keep humans in control while using automation to amplify their capabilities~\cite{shneiderman2022human}. Practical guidelines emphasize making clear what the system can do, showing contextually relevant information, and supporting efficient correction~\cite{amershi2019guidelines}.

A key distinction exists between explanations integrated into the interaction flow and those requiring separate interfaces~\cite{chromik2021taxonomy}. In CLIs and text-based user interfaces (TUIs), explanations are necessarily integrated: the agent's reasoning appears in the same stream as its actions. Combining interpretability with uncertainty awareness enables rapid trust calibration~\cite{tomsett2020rapid}.

\textbf{Implication for other modalities.} In GUI-based agent interaction, actions are visually observable but not easily inspectable or reproducible. The design challenge is to embed transparency into the interaction flow rather than bolt it on separately. Concrete approaches include persistent action logs recording every agent click, inline annotations overlaying the agent's reasoning onto GUI elements, and editable plan representations where the user can modify the agent's next steps before execution.

\section{Property 3: Low Barriers to Human Participation}

\begin{marginfigure}
\begin{tikzpicture}[
    term/.style={fill=ctp-base, text=ctp-text, rounded corners=3pt,
      minimum width=4.4cm, inner sep=4pt, font=\ttfamily\tiny, align=left, text width=4cm}
]
\node[term] (t) {%
\textcolor{ctp-subtext}{\# Traditional CLI:}\\
\textcolor{ctp-red}{\$}\ \textcolor{ctp-subtext}{find . -name "*.py" \textbackslash}\\
\textcolor{ctp-subtext}{\ \ -size +1M -exec ls -lh \{\} +}\\[4pt]
\textcolor{ctp-subtext}{\# With AI agent:}\\
\textcolor{ctp-green}{\$}\ \textcolor{ctp-text}{find all Python files}\\
\textcolor{ctp-green}{\ }\ \textcolor{ctp-text}{larger than 1MB}\\[2pt]
\textcolor{ctp-sky}{Agent:} Running search...\\[2pt]
\textcolor{ctp-teal}{\ \ ./data/analysis.py}\ \textcolor{ctp-subtext}{2.3MB}\\
\textcolor{ctp-teal}{\ \ ./models/train.py}\ \textcolor{ctp-subtext}{1.1MB}\\
\textcolor{ctp-subtext}{Found 2 files.}%
};
\end{tikzpicture}
\caption{Natural language input collapses the Gulf of Execution.}
\label{fig:gulfexec}
\end{marginfigure}

The third property concerns the accessibility of the interface to human users, regardless of their prior expertise with the underlying system.

The historical limitation of CLI was its steep learning curve. Direct manipulation was advocated precisely because users struggled with command recall and syntax~\cite{shneiderman1983direct}, and the Keystroke-Level Model quantified these costs~\cite{card1980keystroke}. Yet even experienced users rarely transitioned from mouse-based to keyboard-based methods: they settled for ``good enough'' rather than investing effort to learn a more efficient alternative~\cite{lane2005hidden, cockburn2014supporting}. With practice, keyboard shortcuts do outperform menu selection, but the crossover point requires substantial investment~\cite{remington2016crossover}.

Natural language input transforms this situation. When an LLM translates natural language into executable commands, the primary barrier to CLI use (command recall and syntactic precision) is eliminated. The Gulf of Execution~\cite{hutchins1985direct} collapses: users specify \textit{what} they want, and the agent determines \textit{how} (Figure~\ref{fig:gulfexec}). The NL2Bash corpus established the feasibility of this translation~\cite{lin2018nl2bash}, and current terminal-based AI tools make it routine. Natural language serves simultaneously as the novice entry point and as the bridge to expert-level capabilities.

The interface design problems identified four decades ago in the direct manipulation literature remain relevant for LLM-powered interfaces~\cite{masson2024directgpt}. The lesson: \textit{any} human-AI-UI modality must address how users with varying expertise levels can participate in the collaboration.

\textbf{Implication for other modalities.} If the agent operates a complex application that the user does not fully understand, oversight becomes difficult regardless of visual clarity. Designing for low barriers means providing natural language channels for directing the agent, progressive disclosure of technical details, and the ability to pause, redirect, or undo agent actions without expertise in the underlying application.

\section{Mixed-Initiative Interaction in the Human-Agent-UI Triangle}

The three properties interact most visibly in mixed-initiative interaction, where both human and agent can take initiative~\cite{horvitz1999principles, skantze2021turnreview}. In terminal-based tools, the CLI prompt provides an unambiguous turn boundary: the user holds initiative by default, and the agent returns it upon completion (Figure~\ref{fig:mixedinitiative}). Each property plays a distinct role. Transparency (P2) lets the human monitor the agent's plan and intervene before irreversible actions. Low barriers (P3) ensure intervention requires only natural language. Representational compatibility (P1) means proposed actions are expressed in the format the human reads. Together, they enable what the automation literature recommends: systems that suggest alternatives while humans approve before execution~\cite{parasuraman2000model}.

\begin{marginfigure}
\begin{tikzpicture}[
    term/.style={fill=ctp-base, text=ctp-text, rounded corners=3pt,
      minimum width=4.4cm, inner sep=4pt, font=\ttfamily\tiny, align=left, text width=4cm}
]
\node[term] (t) {%
\textcolor{ctp-subtext}{\# Human holds initiative:}\\
\textcolor{ctp-green}{\$}\ \textcolor{ctp-text}{add input validation}\\[2pt]
\textcolor{ctp-sky}{Agent:} I'll add validation to\\
\textcolor{ctp-sky}{\ }\ the Flask form.\\
\textcolor{ctp-sky}{\ }\ Plan:\ 1.\ email format\\
\textcolor{ctp-sky}{\ \ \ \ \ \ }\ 2.\ password strength\\
\textcolor{ctp-sky}{\ \ \ \ \ \ }\ 3.\ update pytest suite\\[2pt]
\textcolor{ctp-mauve}{Proceed?}\ \textcolor{ctp-green}{[y/n/edit]}\ \textcolor{ctp-text}{edit}\\[2pt]
\textcolor{ctp-subtext}{\# Human redirects:}\\
\textcolor{ctp-green}{\$}\ \textcolor{ctp-text}{skip password strength,}\\
\textcolor{ctp-green}{\ }\ \textcolor{ctp-text}{add username uniqueness}\\[2pt]
\textcolor{ctp-sky}{Agent:} Updated plan.\\
\textcolor{ctp-sky}{\ }\ 1.\ email format\\
\textcolor{ctp-sky}{\ }\ 2.\ username uniqueness\\
\textcolor{ctp-sky}{\ }\ 3.\ update pytest suite\\
\textcolor{ctp-mauve}{Proceed?}\ \textcolor{ctp-green}{[y/n/edit]}\ \textcolor{ctp-text}{y}%
};
\end{tikzpicture}
\caption{Mixed-initiative interaction: the human redirects the agent's plan mid-task via the CLI prompt.}
\label{fig:mixedinitiative}
\end{marginfigure}

Current CLI agent tools implement this through configurable permission modes, from suggestion-only to fully autonomous execution. The shared codebase provides a common artifact that both parties can reference~\cite{salikutluk2024shared}, and the alternating turn structure enables the bidirectional communication that most co-creative AI systems lack~\cite{rezwana2023cofi}. Human-led interaction produces more diverse outcomes than model-led interaction~\cite{kim2025partnering}, and CLI/TUI tools are inherently human-led: the user initiates each exchange.

For GUI-based agent interaction, these observations suggest that mixed-initiative controls deserve as much attention as perceptual capabilities. An agent that navigates a GUI flawlessly but lacks clear mechanisms for human oversight fails the collaboration requirements identified above. Concretely, GUI agent systems need explicit turn-taking protocols that define when the agent pauses for approval, visible plan representations showing the agent's intended next steps, and low-friction intervention mechanisms that let users redirect mid-task. Terminal-based tools provide all three through the text stream's structure; GUI-based tools must achieve them through deliberate design.

\section{Conclusion}

We have argued that the convergence of terminal-based agentic tools reveals three properties that underlie their effectiveness: representational compatibility between agent and interface, transparency of agent actions within the interaction medium, and low barriers to human participation. These properties are not exclusive to the terminal, but the terminal satisfies them by default whereas other modalities must be deliberately engineered to achieve them. On theoretical grounds drawn from four decades of HCI research, we argue they constitute core design requirements, not optional features, for any Human-AI-UI modality.

For researchers building agents that operate graphical and spatial interfaces, this framing reorients the agenda. The primary bottleneck for GUI agents is not perceptual capability; it is the absence of representational compatibility, transparency, and low barriers to human participation. Enriching graphical interfaces with semantic layers that agents can access directly, embedding action logs and explanations into the interaction flow, and enabling oversight without deep application expertise are all open design challenges that call for the interface design expertise the field has developed over four decades.

The ACI deserves the same rigor and dedicated research attention that the field has devoted to human-facing interface design for the past four decades. The properties identified here provide a starting point: not as a checklist, but as a lens through which to evaluate whether any proposed Human-AI-UI modality can support the collaboration it promises. Our argument draws primarily from software development, where terminal-based agent tools are most mature. A natural empirical test would compare agent interfaces that vary systematically on each property. For instance, a within-subjects study could have participants oversee the same agentic task through a transparent text stream versus an opaque GUI replay, measuring trust calibration, intervention accuracy, and task outcome. Controlled validation across professional contexts beyond software development remains needed. Tasks involving spatial reasoning, visual design, or multimedia content will require modalities that text alone cannot provide, reinforcing that these properties must be engineered into every modality rather than treated as exclusive to the terminal.

\begin{acks}
This work was supported by the Swiss National Science Foundation (SNSF) under Spark grant CRSK-2\_228765.
\end{acks}

\bibliography{references}

\vfill
\noindent\doclicenseThis

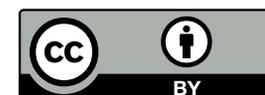
\end{document}